\documentclass[12pt]{article}
\usepackage{graphicx,amsmath,amssymb}
\newcommand{\cinst}[2]{$^{\mathrm{#1)}}$~#2\par}
\newcommand{\crefi}[1]{$^{\mathrm{#1)}}$}
\newcommand{\HRule}{\rule{0.4\linewidth}{0.3mm}}

\newcommand{\kt}{$k_{\rm T}$}

\begin{document}

%%%%%%%%%%%%%%%%%%%%%%%% CERN PREPRINT COVER PAGE
\begingroup
\thispagestyle{empty} \baselineskip=14pt
\parskip 0pt plus 5pt

\begin{center}
{\large EUROPEAN LABORATORY FOR PARTICLE PHYSICS}
\end{center}

\bigskip
\begin{flushright}
CERN--PH--EP\,/\,2010--xxx\\
May 14, 2010
\end{flushright}

\bigskip
\begin{center}
{\Large\bf \boldmath 
Rotation-invariant relations\\[3mm]
in vector meson decays into fermion pairs}

\bigskip\bigskip

Pietro Faccioli\crefi{1},
Carlos Louren\c{c}o\crefi{2},
and Jo\~ao Seixas\crefi{1,3}

\bigskip\bigskip\bigskip
\textbf{Abstract}

\end{center}

\begingroup
\leftskip=0.4cm \rightskip=0.4cm
\parindent=0.pt

  The rotational properties of angular momentum eigenstates imply the
  existence of a frame-independent relation among the parameters of
  the decay distribution of vector mesons into fermions. This relation
  is a generalization of the Lam-Tung identity, a result specific to
  Drell-Yan production in perturbative QCD, here shown to be equivalent to
  the dynamical condition that the dilepton always originates from a
  transversely polarized photon.

\endgroup

%\vspace{1cm}
%\begin{center}
%\emph{Submitted to Phys. Rev. D}
%\end{center}

\vfill
\begin{flushleft}
\HRule\\

\cinst{1} {Laborat\'orio de Instrumenta\c{c}\~ao e F\'{\i}sica Experimental de
  Part\'{\i}culas (LIP),\\ ~~~Lisbon, Portugal} 
\cinst{2} {CERN, Geneva, Switzerland}
\cinst{3} {Physics Department, Instituto Superior T\'ecnico (IST),
  Lisbon, Portugal}
\end{flushleft}
\endgroup

\newpage

%P%\title{Rotation-invariant relations in vector meson decays
%P%into fermions}
%P%
%P%\author{Pietro Faccioli$^a$, Carlos Louren\c{c}o$^b$ and Jo\~ao Seixas$^{a,c}$}
%P%
%P%\affiliation{$^a$LIP, 1000-149 Lisbon, Portugal\\
%P%$^b$CERN, 1211 Geneva, Switzerland\\
%P%$^c$Physics Department, IST, 1049-001 Lisbon, Portugal}
%P%
%P%
%P%\begin{abstract}
%P%
%P%  The rotational properties of angular momentum eigenstates imply the
%P%  existence of a frame-independent relation among the parameters of
%P%  the decay distribution of vector mesons into fermions. This relation
%P%  is a generalization of the Lam-Tung identity, a result specific to
%P%  Drell-Yan production in perturbative QCD, here shown to be equivalent to
%P%  the dynamical condition that the dilepton always originates from a
%P%  transversely polarized photon.
%P%
%P%\end{abstract}
%P%
%P%\pacs{11.80.Cr, 12.38.Qk, 13.20.Gd, 13.85.Qk, 13.88.+e, 14.40.Pq}
%P%
%P%%Keywords:  Lam-Tung relation, polarization, Drell-Yan, quarkonium.
%P%
%P%\maketitle
\sloppy

The partonic cross section for dilepton production in perturbative QCD obeys
the so-called Lam-Tung identity~\cite{bib:LamTung1}, a relation between
the helicity structure functions of the virtual photon or, equivalently,
between the coefficients of the lepton angular distribution measured in the
dilepton rest frame,
$\lambda_\vartheta \, + \, 4 \, \lambda_\varphi \, = \, 1$,
with the $\lambda_\vartheta$ and $\lambda_\varphi$ parameters
introduced later in this Letter.
The Lam-Tung relation (LTR) represents for lepton pair production the
analog of the Callan-Gross relation in deep inelastic scattering, $F_1(x)- 2 x
F_2(x)=F_\mathrm{L}(x)=0$, where the Bjorken scaling functions $F_1$ and $F_2$
are reciprocally connected by the condition that the longitudinal helicity
component $F_\mathrm{L}$ of the massive photon vanishes identically. The
Callan-Gross relation, a consequence of the interaction between the photon
probe and half-integer spin quarks, is not exact, being subject to substantial
$O(\alpha_s)$ corrections due to gluon radiation. 

The theoretical relevance of
the LTR resides in the fact that, although the dilepton production cross
section is substantially modified by QCD corrections, the \textit{relation}
between the different helicity contributions to this cross section remains
unchanged up to $O(\alpha_s^2)$~\cite{bib:LamTung2} and is subject to
relatively small corrections when the subsequent orders in $\alpha_s$ are taken
into account~\cite{bib:BergerQiu}.
In fact, the LTR is such a solid prediction of perturbative QCD that its
violation is a strong signal of non-perturbative effects. 
%Presently, 
Experimentally, the LTR
has been shown to be violated in pion-nucleus collisions~\cite{bib:violation},
raising speculations about the possible quantitative effects of intrinsic
parton \kt~\cite{bib:kt} or of higher twist
contributions~\cite{bib:highertwist}.
Saturation effects are also expected to contribute to a violation of the LTR in
proton-nucleus and deuteron-nucleus collisions at RHIC and at the
LHC~\cite{bib:lowx}.

The distinctive feature of the LTR is that it is independent of the chosen
orientation of the axes of the dilepton rest frame. In this Letter we show that
this property is a completely general consequence of the rotational covariance
of $J = 1$ angular momentum eigenstates. It does not depend, therefore, on the
specific $J = 1$ state considered nor on the production process. This implies
that the decay distribution of any vector state --- including the case of
non-perturbative effects in Drell-Yan production, as well as the case of
quarkonium production --- can be described in terms of a frame-independent
relation analogous to the LTR.

We start by considering the case of a single production ``subprocess'', in
which the vector meson $V$ is always formed as a specific superposition of the
three $J_z$ eigenstates, with eigenvalues $m = +1, -1, 0$, with respect to a
chosen polarization axis $z$:
\begin{equation}
  | V^{(i)} \rangle =  b^{(i)}_{+1} \, |+1\rangle + 
  b^{(i)}_{-1} \, |-1\rangle + b^{(i)}_{0} \, |0\rangle \, . \label{eq:state}
\end{equation}
Assuming helicity conservation at the di-fermion vertex, the angular
distribution of the parity-conserving decay is
%P%\begin{align}
%P%  W^{(i)}(\cos \vartheta, \varphi) \, & \propto \,
%P%  \frac{\mathcal{N}^{(i)}}{(3 + \lambda^{(i)}_{\vartheta})} \,
%P%  (1  + \lambda^{(i)}_{\vartheta} \cos^2 \vartheta \label{eq:ang_distr_subproc} \\
%P%& 
%P%  + \lambda^{(i)}_{\varphi} \sin^2 \vartheta \cos 2 \varphi
%P%  + \lambda^{(i)}_{\vartheta \varphi} \sin 2 \vartheta \cos \varphi
%P%\nonumber \\  & 
%P%  + \lambda^{\bot (i)}_{\varphi} \sin^2 \vartheta \sin 2 \varphi +
%P%  \lambda^{\bot (i)}_{\vartheta \varphi} \sin 2 \vartheta \sin \varphi
%P%  ) \, , \nonumber
%P%\end{align}
%P%
\begin{align}
&  W^{(i)}(\cos \vartheta, \varphi) \, \propto \,
  \frac{\mathcal{N}^{(i)}}{(3 + \lambda^{(i)}_{\vartheta})} \,
  (1  + \lambda^{(i)}_{\vartheta} \cos^2 \vartheta
  + \lambda^{(i)}_{\varphi} \sin^2 \vartheta \cos 2 \varphi
 \label{eq:ang_distr_subproc} \\
& 
  + \lambda^{(i)}_{\vartheta \varphi} \sin 2 \vartheta \cos \varphi
%\nonumber \\  & 
  + \lambda^{\bot (i)}_{\varphi} \sin^2 \vartheta \sin 2 \varphi +
  \lambda^{\bot (i)}_{\vartheta \varphi} \sin 2 \vartheta \sin \varphi
  ) \, , \nonumber
\end{align}
where $\vartheta$ and $\varphi$ are the (polar and azimuthal) angles
formed by the positive fermion with, respectively, the polarization
axis $z$ and the $xz$ plane, and
\begin{align}
\begin{split}
  \mathcal{N}^{(i)} & = |a_0^{(i)}|^2 + |a_{+1}^{(i)}|^2 + |a_{-1}^{(i)}|^2 \, , \\[2mm]
  \lambda^{(i)}_{\vartheta} & = \frac{{\mathcal{N}^{(i)}}-3
    |a_0^{(i)}|^2}{\mathcal{N}^{(i)}+|a_0^{(i)}|^2} \, ,  \label{eq:lambdas_vs_amplitudes}
%P% \\
\quad
%P%
  \lambda^{(i)}_{\varphi} 
%P% & 
=  \frac{ 2 \, \mathrm{Re} (a_{+1}^{(i)*} a_{-1}^{(i)})
}{\mathcal{N}^{(i)}+|a_0^{(i)}|^2} \, , 
%P% \\
\quad
%P%
  \lambda^{(i)}_{\vartheta \varphi}
%P% & 
= \frac{ \sqrt{2} \, \mathrm{Re} [
    a_{0}^{(i)*} ( a_{+1}^{(i)} - a_{-1}^{(i)})]
  }{\mathcal{N}^{(i)}+|a_0^{(i)}|^2} \,
  ,  \\[2mm]
  \lambda^{\bot (i)}_{\varphi}  & = \frac{ - 2 \, \mathrm{Im}
    (a_{+1}^{(i)*} a_{-1}^{(i)}) }{\mathcal{N}^{(i)}+|a_0^{(i)}|^2} \,  , 
%P% \\
\quad
%P%
  \lambda^{\bot (i)}_{\vartheta \varphi} 
%P% &
 = \frac{ - \sqrt{2} \,
    \mathrm{Im} [ a_{0}^{(i)*} ( a_{+1}^{(i)} + a_{-1}^{(i)})]
  }{\mathcal{N}^{(i)}+|a_0^{(i)}|^2} \, ,
\end{split}
\end{align}
with $a_{0}^{(i)}$, $a_{+1}^{(i)}$ and $a_{-1}^{(i)}$ being the partial decay
amplitudes of the three $J_z$ components of the vector state. 

In this Letter we only
consider inclusive production. Therefore, the only possible experimental definition of the
$xz$ plane coincides with the production plane, containing the
directions of the colliding particles and of the decaying particle itself. The
last two terms in Eq.~\ref{eq:ang_distr_subproc} introduce an asymmetry of the
distribution by reflection with respect to the production plane. Such asymmetry
is not forbidden in individual parity-conserving events. In hadronic
collisions, due to the intrinsic parton transverse momenta, for example, the
``natural'' polarization plane does not coincide event-by-event with the
production plane. 
However, the symmetry by reflection must be a property of the
observed \emph{event distribution} when only parity-conserving
processes contribute. 
Indeed, the terms in $\sin^2 \vartheta \sin 2 \varphi$ and $\sin 2 \vartheta
\sin \varphi$ are unobservable, because they vanish on average.

In the presence of $n$ contributing production processes with weights
$f^{(i)}$, the most general \emph{observable} distribution can be
written, therefore, as
\begin{align}
\begin{split}
  W(\cos \vartheta, \varphi) \, & = \, \sum_{i = 1}^{n} f^{(i)}
  W^{(i)}(\cos \vartheta, \varphi) \\[2mm]  & 
  \propto \, \frac{1}{(3 + \lambda_{\vartheta})} \,
  (1 + \lambda_{\vartheta} \cos^2 \vartheta
%P%  \\[2mm]  & 
+ \lambda_{\varphi} \sin^2 \vartheta \cos 2 \varphi +
  \lambda_{\vartheta \varphi} \sin 2 \vartheta \cos \varphi ) \, ,
 \label{eq:ang_distr_general}
\end{split}
\end{align}
where
\begin{align}
\begin{split}
  \lambda_\vartheta \, & = \, \sum_{i = 1}^{n} \frac{f^{(i)}
    \mathcal{N}^{(i)}}{3 + \lambda_{\vartheta}^{(i)}} \,
  \lambda^{(i)}_{\vartheta} \left/ \sum_{i = 1}^{n} \frac{f^{(i)}
      \mathcal{N}^{(i)}}{3 + \lambda_{\vartheta}^{(i)}} \right. \, ,
  \\[2mm]
  \lambda_\varphi \, & = \, \sum_{i = 1}^{n} \frac{f^{(i)}
    \mathcal{N}^{(i)}}{3 + \lambda_{\vartheta}^{(i)}} \,
  \lambda^{(i)}_{\varphi} \left/ \sum_{i = 1}^{n} \frac{f^{(i)}
      \mathcal{N}^{(i)}}{3 + \lambda_{\vartheta}^{(i)}} \right. \, ,
  \\[2mm]
  \lambda_{\vartheta \varphi} \, & = \, \sum_{i = 1}^{n} \frac{f^{(i)}
    \mathcal{N}^{(i)}}{3 + \lambda_{\vartheta}^{(i)}} \,
  \lambda^{(i)}_{\vartheta \varphi} \left/ \sum_{i = 1}^{n}
    \frac{f^{(i)} \mathcal{N}^{(i)}}{3 + \lambda_{\vartheta}^{(i)}}
  \right. \, .
\label{eq:lambdas_general}
\end{split}
\end{align}

Our considerations are based on two propositions concerning the
rotational properties of the generic $J=1$ state defined in
Eq.~\ref{eq:state}.\\
Prop.\ 1: Each amplitude combination $b_{+1}^{(i)}+b_{-1}^{(i)}$ is
invariant by rotation around the $y$ axis.\\
Prop.\ 2: For each subprocess there exists a quantization axis
$z^{(i)*}$ with respect to which $b_0^{(i)*}=0$; if $b_0^{(i)}$,
$b_{+1}^{(i)}$ and $b_{-1}^{(i)}$ are real, $z^{(i)*}$ belongs to the
$xz$ plane.

Proposition~1 follows from the relations among rotation matrix
elements $d_{+1, M}^{1}(\vartheta) + d_{-1, M}^{1}(\vartheta) =
\delta_{|M|,1}$.  When $|V^{(i)} \rangle$ is defined with a real
$b_0^{(i)}$ (always possible), the frame defined in Prop.~2 is reached
through successive rotations by the Euler angles $\varphi^{*}$,
$\vartheta^{*}$ and $-\varphi^{*}$, defined as
\begin{align}
\begin{split}
  \cos \vartheta^{*} & = \frac{R_+ R_- + I_+ I_-}{\sqrt{2 b_0^{(i)2}
      (R_+^2 + I_-^2) +
      (R_+ R_- + I_+ I_-)^2 } } \, , \\
  \cos \varphi^{*} & = \frac{R_+}{\sqrt{R_+^2 + I_-^2}} \, , \quad
  \sin \varphi^{*} = - \frac{I_-}{\sqrt{R_+^2 + I_-^2}} \, ,
\end{split}
\end{align}
where $R_{\pm} = \mathrm{Re}(b_{+1}^{(i)} \pm b_{-1}^{(i)})$ and
$I_{\pm} = \mathrm{Im}(b_{+1}^{(i)} \pm b_{-1}^{(i)})$. If all three
amplitudes are real, then $\varphi^{*} = 0$ and the rotation is around the
$y$ axis.

Proposition~1 and the obvious rotation invariance of $|a_0^{(i)}|^2
+|a_{+1}^{(i)}|^2 + |a_{-1}^{(i)}|^2$ imply that the quantities
\begin{equation}
  \mathcal{F}^{(i)}=\frac{1}{2} \, \frac{|a_{+1}^{(i)}+a_{-1}^{(i)}|^2}{|a_0^{(i)}|^2
    + |a_{+1}^{(i)}|^2 + |a_{-1}^{(i)}|^2} \label{eq:kappa_i}
\end{equation}
(bound between 0 and 1) are independent of the chosen frame. Using
also Eqs.~\ref{eq:lambdas_vs_amplitudes} and \ref{eq:lambdas_general},
we find that the following combination of observable parameters is
frame-independent:
\begin{equation}
  \mathcal{F} \, = \, \frac{ \sum_{i = 1}^{n} f^{(i)}
    \mathcal{N}^{(i)} \mathcal{F}^{(i)}}{\sum_{i =
      1}^{n} f^{(i)} \mathcal{N}^{(i)}}  \, = \, \frac{1 +
    \lambda_\vartheta + 2
    \lambda_\varphi}{3 + \lambda_\vartheta} \, . \label{eq:F}
\end{equation}
Equation~\ref{eq:F} can be written as
\begin{equation}
  (1 - \mathcal{F}) \, (3 + \lambda_\vartheta ) \, = \, 2 \, (1 -
  \lambda_\varphi ) \, , \label{eq:generalized_LamTung}
\end{equation}
an expression formally 
analogous to the LTR~\cite{bib:LamTung1}, which, as mentioned above, accounts for Drell-Yan
production up to first-order QCD modifications, neglecting parton transverse
momenta. 

At this level of description, the topology of each contributing
subprocess (quark-antiquark annihilation without or with single gluon
emission, Compton-like quark-gluon scattering, etc.) is characterized
by one reaction plane, coinciding with the experimental production
plane.  Therefore, for each single subprocess $\lambda^{\bot
  (i)}_{\varphi} = \lambda^{\bot (i)}_{\vartheta \varphi} = 0$,
implying (Eq.~\ref{eq:lambdas_vs_amplitudes}) that the three partial
decay amplitudes (and, thus, the three components of the produced
angular momentum state) can be chosen to be real.  Proposition~2,
together with Eq.~\ref{eq:lambdas_vs_amplitudes}, implies, then, that
the observed dilepton distribution is a convolution of
sub-distributions of the kind
%
%P%\begin{align}
%P%\begin{split}
\begin{equation}
%P% &
  \lambda^{(i)*}_\vartheta = + 1 \, , \quad \lambda^{(i)*}_\varphi =
  2\,\mathcal{F}^{(i)}-1 \, , 
%P% \\ & 
  \lambda^{\bot (i)*}_{\varphi} =
  \lambda^{(i)*}_{\vartheta \varphi} = \lambda^{\bot (i)*}_{\vartheta
    \varphi} = 0 \, , 
\label{eq:elementary_distrib}
\end{equation}
%P%\end{split}
%P%\end{align}
%
each one referred to a specific polarization axis $z^{(i)*}$ belonging to the
production plane. 

The LTR is obtained from Eq.~\ref{eq:generalized_LamTung} in
the special case when the invariants $\mathcal{F}^{(i)}$ (and, thus,
$\mathcal{F}$) are equal to $1/2$. This means, according to
Eq.~\ref{eq:elementary_distrib}, that all competing subprocesses lead to the
same kind of \emph{fully
  transverse, purely polar} decay anisotropy, \emph{with respect to
  possibly different natural axes}. In other words, the
\emph{frame independence} of the LTR is the \emph{kinematic} consequence of the
rotational properties of the $J=1$ angular momentum eigenstates, while its
specific \emph{form} ($\mathcal{F}=1/2$) derives from the \emph{dynamical}
input that all contributing subprocesses produce the dilepton via one transversely
polarized photon.

More generally, the decay of a vector state into fermion pairs in a
given kinematic condition is always described in frame-independent
terms by a specific form of Eq.~\ref{eq:generalized_LamTung}.  The
advantages of this kind of representation, complementary to the
determination of the full angular distribution, are described in
Ref.~\cite{bib:ImprovedPol} for the specific case of quarkonium
polarization studies.

In summary, we have shown that rotational invariance imposes frame-invariant
constraints on the polar and azimuthal anisotropy parameters of the di-fermion
decay distribution of vector mesons. In particular, for any mixture of
production mechanisms in a given kinematic condition there exists a
frame-invariant relation among the angular coefficients, depending on one
calculable parameter, $\mathcal{F}$. The Lam-Tung relation 
corresponds to the special case when all
processes produce the dilepton via one transverse photon and their respective
natural axes belong to the production plane. Any violation of this relation will
continue to be described by a suitably modified \emph{frame-invariant}
relation. The frame-invariant formalism can be extended to the
study of the spin alignment of quarkonium and other vector particles.

\bigskip

P.F.\ and J.S.\ acknowledge support from 
Funda\c{c}\~ao para a Ci\^encia e a Tecnologia,
%%%%%%% (FCT), 
%P%FCT,
%
Portugal, under contracts SFRH/BPD/42343/2007 and CERN/FP/109343/2009.

%%%%%%%%%%%%%%%%%%%%%%%%%%%%%%%%%%%%%%%%%%%%%%%%%%%%%

\end{document}